\documentclass[11pt]{article}
\pdfoutput=1 

\usepackage[T1]{fontenc} % if needed
\def\beq{\begin{equation}}
\def\eeq{\end{equation}}

\newcommand{\bea}{\begin{eqnarray}}
\newcommand{\eea}{\end{eqnarray}}

\usepackage{amsmath} 
\usepackage{multirow}
\usepackage{tabularray}
\usepackage{float}
\usepackage{comment}

\usepackage{microtype}
\usepackage{mathtools}
\usepackage{booktabs}
\usepackage[english]{babel}
\usepackage{amsmath,amssymb,amsbsy,amstext, amsthm, simplewick, amsfonts}
\usepackage{graphicx}
\usepackage[small]{caption}
\usepackage{siunitx}
\usepackage{upgreek}
\usepackage{framed}
\usepackage{wrapfig}
\usepackage{multirow}
\usepackage{bbm}
\usepackage[svgnames,dvipsnames,x11names]{xcolor}
\usepackage[utf8x]{inputenc}
\usepackage{selinput}
\usepackage{bm}
\usepackage{float}
\usepackage[top=2.5cm, bottom=2.5cm, left=3cm, right=3cm]{geometry}
\usepackage{yfonts}
\usepackage{caption}
\usepackage{subcaption}
\usepackage{sidecap}
\usepackage{longtable}
\usepackage{anyfontsize}
\usepackage[hypertexnames = false]{hyperref}
\usepackage{cleveref}

\usepackage{epstopdf}
\usepackage{cancel}
\usepackage{tcolorbox}

\usepackage{pgfornament}

\DeclareFontFamily{OT1}{rsfs10}{}
\DeclareFontShape{OT1}{rsfs10}{m}{n}{ <-> rsfs10 }{}
\DeclareMathAlphabet{\mathscript}{OT1}{rsfs10}{m}{n}

\newcommand{\be}{\begin{equation}}
\newcommand{\ee}{\end{equation}}

\newcommand{\bmat}{\begin{bmatrix}}
\newcommand{\emat}{\end{bmatrix}}

\newcommand{\beqa}{\begin{eqnarray}}
\newcommand{\eeqa}{\end{eqnarray}}
\newcommand{\beqar}{\begin{eqnarray*}}
\newcommand{\eeqar}{\end{eqnarray*}}

\newcommand{\bbibitem}[1]{\bibitem{#1}\marginpar{#1}}

\def\Label#1{\label{#1}%
  \smash{\hbox to0pt{\raise1ex\hbox{\tiny[#1]}\hss}}}
\def\noLabels{\let\Label=\label}
\def\nobbibitem{\let\bbibitem=\bibitem}

\makeatletter
\DeclareRobustCommand{\rcite}[1]{%
  \rcite@aux#1,\@nil{#1}%
}
\def\rcite@aux#1,#2\@nil#3{%
  \if\relax#2\relax
    Ref.~\cite{#3}%
  \else
    Refs.~\cite{#3}%
  \fi
}
\makeatother

\definecolor{greyish}{rgb}{.90,.90,.90}
\definecolor{greyish2}{rgb}{.96,.96,.96}
\usepackage{xcolor,colortbl}
\usepackage{tcolorbox}

\numberwithin{equation}{section}

\begin{document}
\renewcommand{\thefootnote}{\fnsymbol{footnote}}
%~
\vspace{0truecm}
\thispagestyle{empty}

\hfill

\hspace*{0pt}\hfill  UPR-1329-T \\
\hspace*{0pt}\hfill  IFT-UAM/CSIC-24-85 \\

\begin{center}
{\fontsize{21}{18} \bf Kerr Effective Black Hole Geometries}\\[14pt]
{\fontsize{21}{18} \bf  in Supergravity}
\end{center}

\vspace{.15truecm}

\begin{center}
{\fontsize{13}{18}\selectfont
Mirjam Cveti\v{c} $^{\rm a,b}$\footnote{\texttt{cvetic@physics.upenn.edu}} \ ,
Nelson Hern\'andez Rodr\'iguez $^{\rm c}$\footnote{\texttt{nelsonhdez21@gmail.com}} } \ , \\[4.5pt]
{\fontsize{13}{18}\selectfont
Maria J. Rodriguez $^{\rm c,d,e}$\footnote{\texttt{maria.rodriguez@gmail.com}} and 
Oscar Varela\ $^{\rm c,d,e}$\footnote{\texttt{oscar.varela@usu.edu}}
}
\end{center}
\vspace{.4truecm}

\begin{scriptsize}
 \centerline{{\it ${}^{\rm a}$Department of Physics and Astronomy and Department of Mathematics, 
 }}   \centerline{{\it
 University of Pennsylvania, Philadelphia, PA 19104, USA}}

  \vspace{.05cm}

 \centerline{{\it ${}^{\rm b}$Center for Applied Mathematics and Theoretical Physics,
 %}}\centerline{{\it  
 University of Maribor, Maribor, SI2000, Slovenia}}

  \vspace{.05cm}

 \centerline{{\it ${}^{\rm c}$Instituto de F\'\i sica Te\'orica UAM-CSIC,
 %}}\centerline{{\it   C/ Nicolas Cabrera, 13-15
 Madrid, 28049, Spain}}

  \vspace{.05cm}

  \centerline{{\it ${}^{\rm d}$Black Hole Initiative, Harvard University, Cambridge, MA 02138, USA}}

  \vspace{.05cm}

 \centerline{{\it ${}^{\rm e}$Department of Physics, Utah State University, Logan, UT 84322, USA}}

\end{scriptsize}

 \vspace{.25cm}

\vspace{.3cm}
\begin{abstract}
\noindent

We derive the explicit embedding of the effective Kerr spacetimes, which are pertinent to the vanishing of static Love numbers, soft hair descriptions of Kerr black holes, and low-frequency scalar-Kerr scattering amplitudes, as solutions within $N=2$ supergravity. These spacetimes exhibit a hidden $SL(2,R)\times U(1)$ or $SO(4,2)$ symmetry resembling the so called subtracted geometries with $SL(2,R)\times SL(2,R)$ symmetry, which accurately represent the near-horizon geometry of Kerr black holes and, as we will argue most accurately represents the internal structure of the Kerr black hole. To quantify the differences among the effective Kerr spacetimes, we compare their physical quantities, internal structures, and geodesic equations. Although their thermodynamic properties, including entropy, match those of Kerr, our study uncovers significant differences in the interiors of these effective Kerr solutions. A careful examination of the internal structure of the spacetimes highlights the distinctions between various effective Kerr geometries and their quasinormal spectra.

\end{abstract}

\newpage

\setcounter{tocdepth}{2}
\tableofcontents
\newpage
\renewcommand*{\thefootnote}{\arabic{footnote}}
\setcounter{footnote}{0}

\section{Introduction}

\indent Exploring low-frequency physics within black hole backgrounds has proven highly fruitful for the soft hair of Kerr black holes descriptions, low-frequency scalar-Kerr scattering amplitudes and vanishing of the gravitational static Love numbers for rotating black holes. The guiding principle in this approach has been the observation that for certain low frequency regimes, $\omega \ll 1/M $, with $M$ the black hole mass, one can explicitly realize a hidden $SL(2,R)\times SL(2,R)$ symmetry in the massless Klein-Gordon (KG) wave equation for a propagating field. This symmetry does not manifest itself as an isometry of the background metric 

\indent One of the significant effective metrics associated with Kerr black holes is the so-called subtracted geometry \cite{Cvetic:2011dn}. This is specifically connected to the hidden symmetries of Kerr black hole spacetimes via Teukolsky's equation in the theory of gravitational perturbations as employed in \cite{Castro:2010fd}. The key feature is that these geometries provide an approximation in the near-horizon region of a Kerr black hole, maintaining its thermodynamic properties at the outer event horizon and its internal structure.

\indent An additional collection of fundamental geometries pertinent to black holes are the effective near-zone metrics that play a crucial role in describing gravitational tidal deformations, or Love symmetries \cite{Kol_2012,Binnington_2009,LeTiec:2020spy}. These different geometries arise from the ambiguity in how one defines the effective Kerr near zone expansion associated with a freedom to move frequency dependent terms in the Klein-Gordon (KG) equation. For each choice of near zone split, the leading order massless wave equation enjoys a generalized local hidden symmetry \cite{Lowe:2011aa}. Two instances of Kerr Effective Geometry (KEG) approximations have been recently proposed. On the one hand, there is a geometry preserving an $SO(4,2)$ found in \cite{Hui:2022vbh} and, on the other hand, an effective geometry preserving an $SL(2,R)\times U(1)$ as shown in \cite{Charalambous:2021kcz}. These are more mysterious than the $SL(2,R) \times SL(2,R)$; even-though these alternative KEGs preserve the same thermodynamic properties of Kerr, they also realize distinct hidden symmetries of the KG wave equation which could explain the vanishing of the static Love numbers for Kerr black holes. A surprising property of the KEG with  $SO(4,2)$ or $SL(2,R)\times U(1)$ hidden symmetries, is that these do not capture the dynamical Love numbers for Kerr. As recently shown in \cite{Perry:2023wmm}, the KEG with $SL(2,R) \times SL(2,R)$, seems to much closely capture the dynamical Love numbers for Kerr in the low frequency regime.

\indent The lack of uniqueness among these effective geometries poses a significant issue, as each one is associated with distinct hidden symmetries that are presumably unrelated to one another. The goal of the present paper is to quantify the differences among these effective geometries by contrasting their physical quantities and structures. In this context, we will refer to the geometries that lead to local symmetries in the solution space of low frequency massless field configurations as KEGs. 

\indent KEGs are no longer asymptotically flat. Moreover, these metrics with the asymptotic geometry removed in this manner, no longer satisfy Einstein's equations in vacuum. Kerr black holes emerge as solutions within the vacuum Einstein equations, whereas KEGs rely on additional matter content for support. This has been shown in \cite{Cvetic:2012tr} for the subtracted $SL(2,R)\times SL(2,R)$ KEG. In this paper, we will demonstrate explicitly the embedding of more general KEGs as solutions to $N=2$ supergravity in four dimensions.

\indent The structure of this paper is as follows. Section \ref{sec:AC} reviews three distinct effective Kerr metrics, namely, those whose associated leading order KG equation enjoys a $SL(2,R)\times SL(2,R)$, $SO(4,2)$ or $SL(2,R)\times U(1)$ local hidden symmetry. We analyze the structure of these metrics and propose different methods which enable us to calculate the thermodynamics of the KEGs. The field configurations that define the KEGs as solutions of $N = 2$ supergravity can be found in Section \ref{sec:Matter}. In Section \ref{sec:Proof}, additional evidence is presented supporting the proposal that certain KEGs will replicate certain aspects of the physics of Kerr black holes. This section also addresses issues related to the monodromy and internal structure of the spacetimes. Remarkably simple, explicit expressions for the quasinormal mode frequencies of the KEGs are obtained in Section \ref{sec:QNM}. The study of the separability of the wave equation, existence of exact Killing-St\"ackel tensors and geodesic equations is presented in Section \ref{sec:separability}. 
We discuss our results and conclude by presenting some future directions in Section \ref{sec:Disc}. The paper also includes an Appendix that compiles the necessary results for the calculations discussed earlier, presented in a consistent notation.
 We use geometrized units where $G = c = k_B =\hbar = 1$.\\

%%%%%%%%%%%%%%%%%%%%%%SECTION%%%%%%%%%%%%%%%%%%%%%
\section{Kerr Effective Black Hole Geometries}
\label{sec:AC}

In this section we focus on the study of physical properties of the KEGs. The main observation about the KEGs is that these preserve the internal structure of the Kerr black hole as well as the thermodynamic properties.
While Kerr black holes arise as solutions of the vacuum Einstein equations, KEGs are supported by supplementary matter content. These fields supporting such geometries vary as a function of radial distance at spatial infinity, instead of becoming constant. KEGs exhibit distinct asymptotic structures when contrasted with the asymptotically flat and asymptotically AdS scenarios. Moreover, a comprehensive exploration of their conserved charges is lacking, and new insights would provide a starting point for the study of gravity/field theory duality for KEGs. 

\subsection{Definitions}
The spacetime metrics can be treated physically as black holes confined in a box. The KEG geometry is
\begin{equation}\label{ACmetric}
ds_{KEG}^{2}=-\Delta_{0}^{-1/2} G_{0} \, (dt+A_{0}\,d\phi)^{2} + \Delta_{0}^{1/2}\left(\frac{dr^{2}}{\Delta}+d\theta^{2} + \frac{\Delta}{G_{0}}\sin^{2}\theta\, d\phi^{2}\right)
\end{equation}
with $\Delta_0$ a function of $r$ and,
\bea
\left. A_0\,\, \right|_{r=r_+}=- (r_+^2+a^2)\,,\qquad \left. G_0\,\, \right|_{r=r_+}= -a^2\sin^2\theta\,,
\eea
such that effectively the near black hole horizon Kerr region is preserved. The corresponding asymptotic form of the (static) geometry reads
\begin{equation}\label{ACmetric1}
ds_{AsympKEG}^{2}=-Y^{2p} dt^2+B^{2}\frac{dY^2}{Y^{2s}}+Y^{2q}(d\theta^2+\sin^2\theta d\phi^2)\,,
\end{equation}
where $(p,q,s,B)$ are constants; in particular KEGs are not asymptotically flat implying $p,s\ne0$, and $q,B\ne1$.

For Kerr black holes in standard Boyer-Lindquist coordinates, the warp factor $\Delta_0$, the function $\Delta$ and angular potential 
$A_0$ are given by
\bea
& G^{Kerr}_{0}  = \Delta -a^{2}\sin^{2}\theta \,, \qquad A^{Kerr}_{0}  = \frac{2 M a \, r \sin^{2}\theta}{G^{Kerr}_{0}}\,, \\
&\Delta^{Kerr}_{0} = (r^2+ a^2\cos^{2}\theta)^2\,, \qquad \Delta^{Kerr}  =r^{2}+a^{2}-2M r \, .
\eea
However, asymptotically the geometry is flat, taking the form \eqref{ACmetric1} with $s,p=0$ and $q,B=1$.

The main observation leading to the KEG proposals is that the thermodynamic properties of the Kerr black hole solution is completely independent of the warping factor $\Delta_0$. A fact that suggests that modifying $\Delta^{Kerr}_0$ in \eqref{ACmetric1} preserves the internal near structure and retains the same thermodynamic properties of Kerr black holes.

Concrete KEGs of the form (\ref{ACmetric}) arise for different choices of the metric functions. All three of them that we will consider in this paper lead to a massless wave equation,
\bea\label{Laplacian}
\Delta^{-1/2}_0\left[
\partial_r \Delta \partial_r + \frac{1}{\sin\theta} \partial_\theta
\sin\theta \partial_\theta - \frac{\Delta_0} {G_{0}}\partial^2_t
+\frac{G_{0}}{\Delta\sin^2\theta} (\partial_\phi - A_{0}\partial_t)^2
\right] \psi =0 \ , 
\eea
with at least $SL(2, R)$ hidden symmetry \cite{Lowe:2011aa}. The first such geometry allows for modifications with respect to Kerr only in the warping factor $\Delta_0$, leaving the angular potential intact, $A_0=A^{Kerr}_0$. This geometry is the subtracted geometry, also called black hole a box, introduced in \cite{Cvetic:2011dn,Cvetic:2012tr}. The subtracted geometry is asymptotically conical with a Lifshitz type symmetry (a diffeomorphism under which the pull-back metric goes into a constant multiple of itself), with time and radial
distance scaling differently. Two other KEG geometries arise that alter both the warping factor $\Delta_0$ in \eqref{ACmetric} and the angular potential $A_0$ with respect to Kerr. These lead to wave equations (\ref{Laplacian}) that display hidden $SL(2,R)\times U(1)$ \cite{Charalambous:2021kcz} and $SO(4,2)$ symmetries \cite{Hui:2022vbh}, respectively, and are relevant in the context of tidal Love numbers. These have different asymptotics. It is worth emphasizing that all these hidden symmetries do not correspond to isometries of the background metric (\ref{ACmetric}) but rather manifest themselves in the massless scalar wave equation (\ref{Laplacian}). The next subsections review these KEGs and compare the corresponding thermodynamic properties in each case.

\subsection{Effective \texorpdfstring{$SL(2,R)\times SL(2,R)$}{SL(2,R) X SL(2,R)} Black Hole Geometries}
\label{subsec1}

Let us review the effective geometry that generates the near region in \cite{Castro:2010fd} that captures the $SL(2, R) \times SL(2, R)$ symmetry of the near-region
scalar field equation. In terms of the $(t,r,\theta,\phi)$ coordinates, the corresponding effective metric is \eqref{ACmetric}
with
\bea\label{SL2Reqs}
& G_{0}  =\nonumber \Delta -a^{2}\sin^{2}\theta \,, \qquad A_{0}  = \frac{2 m a \, r \sin^{2}\theta}{G_{0}}\,, \\
&\Delta_{0} = 4 m^2(2m r- a^2\cos^{2}\theta)\,, \qquad \Delta  =r^{2}+a^{2}-2m r \, .
\eea
This metric coincides with the one presented in \cite{Cvetic:2014nta} in the chargeless limit. At the specific points where $\Delta_0=0$, the Ricci scalar diverges, suggesting a singularity in the curvature of the metric. This curvature singularity lies behind the black hole horizons.

It is interesting to study the behavior of these asymptotically conical metrics as we approach the limit where $r$ tends to infinity. At leading order, the following expression emerges,

\begin{align}
ds^{2} \nonumber (Asymp)_{SL(2,R)\times SL(2,R)}\sim  & -\sqrt{\frac{r^{3}}{8m^{3}}}dt^2 -2 a \sin^2 \theta \sqrt{\frac{r}{2m}}dtd\phi \\
 & +\sqrt{8m^{3}r}\,\left(\frac{dr^{2}}{r^2} + d\theta^{2} -\sin^{2}\theta d\phi^{2}\right).
\end{align}

A coordinate transformation $\phi\rightarrow \varphi 
+ A t$ where $A=a/(4m^2)$ simplifies the metric corresponding to the asymptotic form \eqref{ACmetric1}. In this case, $p=3,q=1,s=0,B=4$ and $Y=(8m^3 \, r)^{1/4}$ as shown in \cite{Cvetic:2014nta}.

\subsection{Effective \texorpdfstring{$SL(2,R)\times U(1)$}{SL(2,R) X U(1)} Black Hole Geometries}
\label{subsec2}

One of the metrics that has garnered increasing interest for its fundamental role in describing Love symmetries is the one found in \cite{Charalambous:2021kcz}. In this particular case, the hidden ``Love'' symmetry possessed by this metric corresponds to $SL(2,R)\times U(1)$. 
Once again, the line element is defined by \eqref{ACmetric}, but in this case the functions are 
\begin{align} \label{eq:LoveFuns}
G_{0} & = \nonumber\Delta -a^{2}\sin^{2}\theta, \qquad A_{0} = \frac{a \sin^{2}\theta}{G_{0}}(r_{+}^{2}+a^{2}+\beta(r-r_{+})), \\
\Delta_{0} & = (r_{+}^{2}+a^{2})^{2}(1+\beta^{2}\Omega^{2}\sin^{2}\theta),\qquad \Delta =r^{2}+a^{2}-2mr,
\end{align}
where 
\bea
\beta = \frac{1}{2\pi T}\,,\qquad \text{and} \qquad  T\equiv \frac{\kappa_+}{2\pi}= \frac{r_+-r_-}{8\pi M r_+}\,
\eea
is the temperature of the black hole. The corresponding Ricci scalar exhibits a divergence at $\Delta_0=0$, indicating a curvature singularity in the metric.

The behavior at infinity of this metrics differs completely from the other geometries. As we approach the limit of $r$, the metric of the black hole behaves as 
\begin{equation}
     ds^{2}(Asymp)_{SL(2,R)\times U(1)}\sim  -\frac{r^{2}}{\Gamma }\left(dt+\frac{a\beta \sin^{2}\theta d\phi}{r}\right)^{2}+ \Gamma\left(\frac{dr^{2}}{r^{2}} +  d\theta^{2} + \sin^{2}\theta  d\phi^{2}\right)
\end{equation}
where the function $\Gamma$ depends on $\theta$ as
\begin{equation}
\Gamma= (a^{2}+r_{+}^{2})\sqrt{1+\beta^{2} \Omega^{2}\sin^{2}\theta}.
\end{equation}

Note that only in the static case, for $a=0$, the geometry is of the form \eqref{ACmetric1}.
In this static limit the asymptotic geometry has a scaling symmetry $ds^{2}_{SL(2,R)\times U(1)}\rightarrow \epsilon^0  \,ds^{2}_{SL(2,R)\times U(1)}$ implemented by taking $r\rightarrow \epsilon r$ and $t\rightarrow \epsilon^{-1} t$.

\subsection{Effective \texorpdfstring{$SO(4,2)$}{SL(2,R) X U(1)} Black Hole Geometries}
\label{subsec3}

 Finally, the last metric we will consider was presented in \cite{Hui:2022vbh}, and has a ``Starobinsky'' hidden symmetry $SO(4,2)$. Similar to the previous cases, the line element will be \eqref{ACmetric}; however, the functions will depend on how we perform our effective near-zone approximation. Specifically, we consider

\begin{align}\label{equs}
G_{0} & = \nonumber \Delta -a^{2}\sin^{2}\theta, \qquad
A_{0} = \frac{a \sin^{2}\theta}{G_{0}}(r_{+}^{2}+a^{2}), \\
\Delta_{0} & = (r_{+}^{2}+a^{2})^{2},\qquad
\Delta =r^{2}+a^{2}-2mr.
\end{align}
The metric is a conformally flat spacetime, has vanishing Ricci scalar (though not vanishing Ricci tensor), and represents a solution supported by external fields described in section \ref{sec:Matter}.

The asymptotic behavior of the metric in the limit $r \rightarrow\infty$ is
\begin{align}
ds^{2}\nonumber (Asymp)_{SO(4,2)}\sim & 
-\frac{r^2}{(r^2_++a^2)} dt^2- 2 a\sin^2\theta dt\, d\phi\\
& +(r^2_++a^2)\left(\frac{dr^{2}}{r^{2}} +  d\theta^{2} + \sin^{2}\theta  d\phi^{2}\right)
\end{align}
A coordinate transformation $\phi\rightarrow \varphi 
+ A t$ with $A=-\frac{a}{r_+^2+a^2}$ simplifies the metric corresponding to the asymptotic form \eqref{ACmetric1} with the choice of parameters $s=p=1$, $B=1$ and $q=0$, which locally is equivalent to $\text{AdS}_{2} \times S^{2}$.

\subsection{Thermodynamic properties}
\label{subsec:thermo}

We have determined explicitly the thermodynamic quantities of the KEGs of the previous subsections.  The entropy, temperature and angular velocity of the original Kerr black hole remain unchanged. Explicitly, the entropy of the KEGs \eqref{ACmetric}  is given by
\begin{equation}
S= \frac{A}{4}=\left.\int_{0}^{2\pi}\int_{0}^{\pi}\sqrt{g_{\theta\theta}g_{\phi\phi}} \,\,\right|_{r=r_{+}}d\theta d\phi= \pi\, (r_{+}^{2}+a^{2}).
\end{equation}
The temperature and angular velocity on the event horizon are respectively defined as
\begin{equation}
T 
\equiv \frac{\kappa}{2\pi}=
\left.\frac{(N^{2})'}{4 \pi  \sqrt{N^{2}g_{rr}}}\right|_{r=r_{+}}=\frac{\Delta^{'}(r_{+})}{4\pi (r_{+}^{2}+a^{2})},
\end{equation}
where $\kappa$ is the surface gravity, $N$ is the lapse function, and $\Delta^{'}(r_{+})$ is the derivative of function $\Delta$ evaluated at the outer horizon, and
\begin{equation}
\Omega=-\left. \frac{g_{t\phi}}{g_{\phi\phi}} \,\, \right|_{r=r_{+}}=\frac{a}{r_{+}^{2}+a^{2}}.
\end{equation}
The black hole outer and inner horizons ($\Delta=0$) remain at
\begin{equation}
r_{\pm}=m\pm\sqrt{m^{2}-a^{2}}\,.
\end{equation}
Likewise, the ergosphere defined by $g_{tt}=0$, is only determined by $G_0$ for this type of black holes. In this sense, for all KEGs considered here, the black hole's internal structure is preserved and is determined by the following expression
\begin{equation}
r^{erg}_{\pm}=m\pm\sqrt{m^{2}-a^{2}\cos^{2}\theta}\,.
\end{equation}
The definitions of mass and angular momentum are significantly influenced by the asymptotic properties of the curved geometry. Since the KEGs do not exhibit asymptotic flatness, calculating these conserved quantities becomes quite challenging.

The corresponding intensive quantities are defined at the
inner Cauchy horizon for the KEG. References \cite{Cvetic:2010mn,Castro:2012av} studied these relations for black holes, and in particular for a Kerr black hole the angular velocity evaluated at the Cauchy horizon was found in \cite{Curir-Francaviglia-1979}
\bea
\Omega^{Kerr}_-=-\left. \frac{g_{t\phi}}{g_{\phi\phi}} \,\, \right|_{r=r_{-}}=\frac{a}{r_{-}^{2}+a^{2}}
\eea
Extending this definition for other geometries we can find the angular velocity at the inner Cauchy horizon for the KEGs. We find
\bea
&&\Omega_-^{SL(2,R)\times SL(2,R)}=\frac{a}{r_{-}^{2}+a^{2}}=\Omega^{Kerr}_-,\\
&&\Omega_-^{SO(4,2)}=-\Omega_-^{SL(2,R)\times U(1)}=\frac{a}{r_{+}^{2}+a^{2}}=\Omega.
\eea

As for the temperature of the Cauchy horizon, the Kerr black hole has 
\begin{equation}
 T_{-}^{Kerr} 
 \equiv -\frac{\kappa_-}{2\pi}
 = -\left.\frac{(N^{2})'}{4 \pi  \sqrt{N^{2}g_{rr}}}\right|_{r=r_{-}}= - \frac{r_{+}-r_{-}}{8 \pi M r_{-}} .
    \end{equation}
For the KEGs we find
\begin{align}
    T_{-}^{SL(2,R)\times SL(2,R)}& = -\frac{r_{+}-r_{-}}{8 \pi M r_{-}}=T_{-}^{Kerr} \\
    T_{-}^{SL(2,R)\times U(1)} & = -\frac{r_{+}-r_{-}}{8 \pi M r_{+}}=-T_{-}^{SO(4,2)} 
\end{align}

\section{Matter Supporting the KEGs}
\label{sec:Matter}

The KEGs under consideration do not satisfy Einstein's equations of motion in vacuum. In this section we explicitly identify the matter configurations that support these geometries.

\subsection{The supergravity model} \label{sec:sugramodel}

Any geometry is a solution to Einstein's equations if the energy momentum tensor is chosen as the Einstein tensor of the geometry. A standard criterion for the physical viability of that matter content involves specifying suitable energy conditions for the stress tensor. For our KEGs, all energy conditions are satisfied. In fact, all KEGs arise as solutions of $N=2$ supergravity coupled to three vector multiplets, namely, the so-called STU model. This was shown in \cite{Cvetic:2012tr} for the subtracted geometries with hidden $SL(2,R)\times SL(2,R)$ symmetry. Here, we will extend that proof to the other KEGs under consideration.

The bosonic Lagrangian of the STU model is \cite{Chong:2004na} 
\bea
{\cal L}_4 &=& R\, {*{\bf 1}} - \frac{1}{2} {*d\varphi_i}\wedge d\varphi_i 
   - \frac{1}{2} e^{2\varphi_i}\, {*d\chi_i}\wedge d\chi_i - \frac{1}{2} e^{-\varphi_1}\,
( e^{\varphi_2-\varphi_3}\, {*  F_{(2)1}}\wedge   F_{(2) 1}\nonumber\cr
 &+& e^{\varphi_2+\varphi_3}\, {*   F_{(2)2}}\wedge   F_{(2) 2}\nonumber
  + e^{-\varphi_2 + \varphi_3}\, {*  {\cal F}^1_{(2)} }\wedge   {\cal F}^1_{(2)} + 
     e^{-\varphi_2 -\varphi_3}\, {* {\cal F}^2_{(2)}}\wedge   {\cal F}^2_{(2)})\nonumber\\
&-& \chi_1\, (  F_{(2)1}\wedge  {\cal F}^1_{(2)} + 
                   F_{(2) 2}\wedge  {\cal F}^2_{(2)})\, .
\label{d4lag}
\eea
%%%%%
The index $ i = 1, 2 ,3 $ labels the vector multiplets. The dilaton and axion scalar fields are respectively denoted as $\varphi_{i}$ and $\chi_{i}$. Finally, the model also contains four $U(1)$ gauge fields, $A_{(1)1}$, $A_{(1) 2}$, ${\cal A}^1_{(1)}$ and ${\cal A}^2_{(1)}$, with field strengths defined as 
%%%%%
\bea
  F_{(2)1} &=& d   A_{(1)1} - \chi_2\, d {\cal A}^2_{(1)} , \nonumber\\
  F_{(2) 2} &=& d  A_{(1) 2} + \chi_2\, d {\cal A}^1_{(1)} - \chi_3\, d   A_{(1) 1} +\chi_2\, \chi_3\, d  {\cal A}^2_{(1)} , \nonumber\\
  {\cal F}^1_{(2)} &=& d  {\cal A}^1_{(1)} + \chi_3\, d  {\cal A}^2_{(1)} , \nonumber\\
  {\cal F}^2_{(2)} &=& d  {\cal A}^2_{(1)} .
\eea

The $SL(2,R)\times SL(2,R)$ subtracted geometry, \eqref{ACmetric} with \eqref{SL2Reqs}, was shown in \cite{Cvetic:2012tr} to arise as a solution of the STU model (\ref{d4lag}). This was done through a rescaling limit that we briefly review in Appendix \ref{SL2R}. As we will show next, the other KEGs are also solutions of (\ref{d4lag}). The starting point is the asymptotically flat black hole solution \cite{Cvetic:1996kv,Chong:2004na} (see also \cite{Chow:2014cca}) 
\bea
d{ s}^2_4 & = -{  \Delta}^{-1/2}_0 { G} ( d{ \bar{t}}+{ {\cal  A}})^2 + { \Delta}^{1/2}_0 
\left( \frac{d \bar{r}^2}{X}+ d\theta^2 + \frac{X}{G} \sin^2\theta d\bar{\phi}^2\right)~,\label{metricg4d}
\eea
where 
\bea
{ X} & =& { \bar{r}}^2 - 2{ \bar{m}}{ \bar{r}} + { \bar{a}}^2~,\cr
{ G} & = &{ \bar{r}}^2 - 2{  \bar{m}}{ \bar{r}} + { \bar{a}}^2 \cos^2\theta ~, \cr
{ {\cal A}} & =& \frac{2\bar{m}\bar{a}\sin^2\theta}{G}
\left[ ({ \Pi_c} - { \Pi_s}){  \bar{r}} + 2{ \bar{m}}{ \Pi}_s\right] d\bar{\phi}~,
\eea
and
\bea \label{otros}
{{\Delta}}_0 =&& \prod_{I=1}^4 ({ \bar{r}} + 2{ \bar{m}}\sinh^2 { \bar{\delta}}_I)
+ 2 { \bar{a}}^2 \cos^2\theta [{ \bar{r}}^2 + { \bar{m}}{ \bar{r}}\sum_{I=1}^4\sinh^2{ \bar{\delta}_I}
+\,  4{ \bar{m}}^2 ({ \Pi}_c - { \Pi}_s){ \Pi}_s  \cr && -  2{ \bar{m}}^2 \sum_{I<J<K}
\sinh^2 { \bar{\delta}}_I\sinh^2 { \bar{\delta}}_J\sinh^2 { \bar{\delta}}_K]
+ { \bar{a}}^4 \cos^4\theta~.   
\eea
We are employing the following abbreviations: 
\beq
{ \Pi}_c \equiv \prod_{I=1}^4\cosh{ \bar{\delta}}_I 
~,~~~ { \Pi}_s \equiv  \prod_{I=1}^4 \sinh{ \bar{\delta}}_I~.
\eeq
Please refer to appendix \ref{app:A} for the explicit expressions of the scalars and gauge fields that support the black hole geometry (\ref{metricg4d}).

Let us first implement the scaling limit in the static $SL(2,R)\times U(1)$ and $SO(4,2)$ cases, before moving to the rotating cases. 

\subsection{ Static case}

Consider first the static, $a=0$, limit, of the black hole (\ref{metricg4d}). This metric is a solution to the field equations that follow from (\ref{d4lag}), together with the scalars and vectors specified in appendix \ref{app:A} with $a=0$. For example, the scalar fields take the form
\bea
\label{fields}
&&\chi_i=0 \quad ( i=1,2,3) \nonumber \\
&&e^{\varphi_1} = \left[\frac{(\bar{r}+2\bar{m}\sinh^2\bar{\delta}_1)(\bar{r}+2\bar{m}\sinh^2\bar{\delta}_3)}{(\bar{r}+2\bar{m}\sinh^2\bar{\delta}_2)(\bar{r}+2\bar{m}\sinh^2\bar{\delta}_4)}\right]^{\frac{1}{2}}\,  e^{\varphi_2} \nonumber = \left[\frac{(\bar{r}+2\bar{m}\sinh^2\bar{\delta}_2)(\bar{r}+2\bar{m}\sinh^2\bar{\delta}_3)}{(\bar{r}+2\bar{m}\sinh^2\bar{\delta}_1)(\bar{r}+2\bar{m}\sinh^2\bar{\delta}_4)}\right]^{\frac{1}{2}}\\
&&
e^{\varphi_3} = \left[\frac{(\bar{r}+2\bar{m}\sinh^2\bar{\delta}_1)(\bar{r}+2\bar{m}\sinh^2\bar{\delta}_2)}{(\bar{r}+2\bar{m}\sinh^2\bar{\delta}_3)(\bar{r}+2\bar{m}\sinh^2\bar{\delta}_4)}\right]^{\frac{1}{2}}.
\eea
We have put bars over all variables in the original $N = 2$ supergravity solution. Next, we choose the charge parameters to be all equal, as $\bar{\delta}_1=\bar{\delta}_2 =\bar{\delta}_3=\bar{\delta}_4\equiv \bar{\delta}$. This reduces the solution to the the Reissner-Nordström black hole, as all scalars become zero and the gauge fields become related to a single Einstein-Maxwell gauge field or its Hodge dual. Finally, we rescale the original variables as  
\bea
\label{scaling}
\nonumber\bar{r}= r \epsilon \,,\qquad \bar{t} =t \epsilon^{-1}\,,\qquad  \bar{\phi}= \phi \,,\qquad \bar{m}=m \,\epsilon\,,\\
2 \,\bar{m}\sinh^2\bar{\delta} \equiv Q \,,\qquad  \sinh^2\bar{\delta} \equiv \epsilon^{-1} \sinh^2{\delta} \, ,
\eea
and take the $\epsilon \rightarrow 0$ limit. This limit is finite in our new cases, in contrast to the formally infinite limiting case considered in \cite{Cvetic:2012tr}. 

In the limit, the metric becomes the (static) KEG geometry \eqref{ACmetric}, with the functions defined in either Section \ref{subsec2} or Section \ref{subsec3}. The scalars (\ref{fields}) vanish (even before taking the limit, as noted above)
\bea
\chi_1=\chi_2=\chi_3=0,\qquad {\varphi_1}={\varphi_2}={\varphi_3}= 0\,,
\eea
the electric gauge fields become 
\begin{align} \label{eq:AdS2S2ElecVec}
    A_{(1)2}^{Star}=(\mathcal{A}_{(1)}^{2})^{Star}= \frac{(r-m)}{2 \, m \sinh^{2}\delta}\,dt 
\end{align}
and the magnetic ones 
\begin{align} \label{eq:AdS2S2MagVec}
    A_{(1)1}^{Star}= \mathcal{A}_{(1)1}^{Star}=-2 \, m \cos\theta \sinh^2\delta \,d\phi 
\end{align}

The vectors (\ref{eq:AdS2S2MagVec}) are dual to (\ref{eq:AdS2S2ElecVec}) and have constant electric field strength. The limiting configuration turns out to be the Bertotti-Robinson solution AdS$_2 \times S^2$ of $N=2$ pure supergravity, namely, Einstein-Maxwell theory.

The realization of these geometric configurations as solutions within the same theory as the original black holes implies a more immediate connection between these theories. We show that the KEG can be obtained by a solution generating technique (within the STU-model) emerging as a distinct Harrison transformation within the framework of Dilaton-Maxwell-Einstein gravity theory in Appendix \ref{app:B}. Recognizing the transformation in the non-rotational scenario, as we have, might also offer a practical approach for extending these ideas to cases involving rotations following \cite{Virmani:2012kw,Cvetic:2013cja,Sahay:2013xda}.

\subsection{Rotating case}

We now derive the rotating KEGs as solutions of the $N=2$ supergravity model of section \ref{sec:sugramodel}.  
The metric is \eqref{ACmetric} above, but now we write all the quantities in the original geometry with tildes, for convenience. We also set
\begin{equation} \label{eq:deltachoice}
\tilde{\delta}_1 = \tilde{\delta}_2 = \tilde{\delta}_3 \equiv \tilde{\delta} \; , \quad  \mathrm{and} \quad  \tilde{\delta}_4 \equiv \tilde{\delta}_0 \; .
\end{equation}
It is useful to note that, after these changes, the scalar fields take on the form 
\begin{align}
&\chi_1=\chi_2=\chi_3= \frac{2{\bar m}\, {\bar a}\cos\theta\, \cosh{\bar \delta} \sinh {\bar \delta}(\cosh{\bar \delta}\sinh{\bar \delta}_0-\sinh {\bar \delta}\cosh {\bar \delta} _0)}{({\bar r}+2{\bar m}\sinh^2 {\bar \delta})^2+ {\bar a}^2\cos^2\theta},\\
&e^{\varphi_1} =e^{\varphi_2} =e^{\varphi_3} = \frac{({\bar r}+2{\bar m}\sinh^2 {\bar \delta})^2+ {\bar a}^2\cos^2\theta}{{\bar {\Delta}_0}^{1/2}}\,
\end{align}
The corresponding gauge potentials follow from the expressions given in Appendix \ref{app:A}.

\subsection*{Starobinski Scaling}

Let us first the rescaling corresponding to the $SO(4,2)$ Starobinski scenario. With the choice of charges (\ref{eq:deltachoice}), we next 
adopt the rescaling
\bea
\label{scaling1}
\nonumber\bar{r}= r \epsilon \,,\qquad \bar{t} =t \epsilon^{-1}\,,\qquad  \bar{\phi}= \epsilon \phi \,,\qquad \bar{m}=m \,\epsilon\,,\\
 \bar{a}=a \,\epsilon\,, \qquad
 \sinh\bar{\delta} = \sinh \bar{\delta_0} \equiv   \epsilon^{-1/2} \sinh{\delta} 
\eea
and then send $\epsilon \rightarrow 0$. In the limit, the metric \eqref{ACmetric} with (\ref{equs}) is recovered, provided one identifies $\sinh^4{\delta} \equiv r_+/(r_++r_-)$. This configuration has again vanishing scalar fields,
\bea
\chi_1=\chi_2=\chi_3=0,\qquad {\varphi_1}={\varphi_2}={\varphi_3}= 0\,,
\eea
but the gauge fields become 
\begin{align}
    A_{(1)2}^{Star}=(\mathcal{A}_{(1)}^{2})^{Star}= \frac{(r-m)}{2 m \sinh^{2}\delta}\,dt 
\end{align}

and
\begin{align}
    A_{(1)1}^{Star}= \mathcal{A}_{(1)1}^{Star}=\frac{a\, \cos\theta }{2 m \sinh ^2\delta} \, dt -2 m \cos\theta \sinh^2\delta \,d\phi \; .
\end{align}

We have explicitly verified that this configuration solves the field equations that derive from the Lagrangian (\ref{d4lag}).

Both the original $N=2$ Supergravity solution and the $SO(4,2)$ KEG are parameterized respectively by charge variables $\bar{\delta}$ and $\delta$. We find that there is an interesting enhancement of symmetry, that resembles the Kerr-Newman black hole solution,
as all four charges are identical.

\subsection*{Love Scaling}

In order to recover the $SL(2,R) \times U(1)$ geometry, we again set the charge parameters as in \ref{eq:deltachoice}, but now rescale the original quantities as 
\bea
\label{scaling2}
 &\bar{r}= r \epsilon \,,\qquad \bar{t} =t \epsilon^{-1}\,,\qquad  \bar{m}=m \,\epsilon\,,\\
& \bar{a}=a \,\epsilon\,, \qquad
 \sinh\bar{\delta} = -\sinh \bar{\delta_0} \equiv  \epsilon^{-1/2} \sinh{\delta} . \nonumber
\eea

We recover the Love geometry, (\ref{ACmetric}) with (\ref{eq:LoveFuns}), in the limit $\epsilon \rightarrow 0$ upon  setting $\sinh^4{\delta}=r_+/(r_+-r_-)$. The scalar fields become 
\bea
\chi_1=-\chi_2=\chi_3=-\frac{a}{m} \cos\theta\,,\\
e^{\varphi_1}=e^{\varphi_2}=e^{\varphi_3}=\frac{m}{\sqrt{m^2-a^2\cos^2\theta}}\,.
\eea
Finally, the gauge potentials are given by 

\begin{align}
    A_{(1)2}^{Love}=-(\mathcal{A}_{(1)}^{2})^{Love}= \frac{m(m-r)\hspace{0.1cm}}{2 \sinh^{2}\delta\, (m^2-a^{2}\cos^2\theta)} \, dt -\frac{2 a m^2  \,\sinh ^2\delta \sin^2\theta}{(m^2-a^2 \cos ^2\theta)} \, d\phi
\end{align}
and
 \begin{align}
    A_{(1)1}^{Love}= (\mathcal{A}_{(1)}^{1})^{Love}=-\frac{a(m-r) \cos\theta\hspace{0.1cm}}{2 \sinh^{2}\delta\, (m^2-a^{2}\cos^2\theta)} \, dt -\frac{2 m (m^2-a^2)  \,\sinh ^2\delta \cos\theta}{(m^2-a^2 \cos ^2\theta)} \, d\phi .
\end{align}
We have again verified that this configuration solves the equations of motion that follow from the Lagrangian (\ref{d4lag}).

We can now write the electric charges
\bea
\label{eq:ChargeLove}
Q=-\mathcal{Q} =\frac{1}{2} m \,\sinh^2\delta\,.
\eea
An important point to notice here is that the dilatons and axions have a $\theta$-dependence.
Interestingly, as for the subtracted KEG in \cite{Cvetic:2012tr}, the gauge fields supporting such KEG geometries vary as a function of radial distance at infinity. This fact will become relevant in the definitions of the asymptotic charges. 

Remarkably, we have successfully recovered the previously established scaling limit \eqref{scaling} when the spin parameter vanishes -- it is  static, and the spin is set to zero $a=0$. This not only underscores the robustness of our rescaling but also offers valuable insights into the behavior of the functions under specific scaling conditions.

%%%%%%%%%%%%%%%%%%% SECTION %%%%%%%%%%%%%%%%
\section{Monodromies and Internal Structure}
\label{sec:Proof}

If one is interested in Kerr black holes far from extremality, an interesting option to consider are KEGs. All three KEGs in this paper are contained in the one-parameter family generating local symmetries in the solution space of low frequency massless field perturbations in the general Kerr
geometry \cite{Lowe:2011aa}. These possess an $SL(2,R)\times SL(2,R)$ symmetry, except at two special points corresponding to a reduction of the algebra to an $SL(2,R)$ symmetry of the Schwarzschild background (with $a=0$). One corresponding to the KEG with $SL(2, R) \times U (1) $ hidden symmetry in Section \ref{subsec2} and  $SO(4,2)$ in Section \ref{subsec3}. A detailed analysis of the wave equation generated by the KEG in Section \ref{subsec1} can be found in \cite{Castro:2012av}. The radial second order differential equation reduces to  
\bea\label{eq:KG1}
\Bigg[\partial_r \left(\Delta_r  \partial_r \right)+  \frac{\left(2M\omega r_+  - a m\right)^2}{(r-r_+)(r_+ - r_-)}  - \frac{\left(2M\omega r_-  - a m\right)^2}{(r-r_-)(r_+ - r_-)}  -  \hat{K}_{\ell, s} \Bigg] \hat{R}_s = 0
\eea
 where the $\hat{K}_{\ell, s}$ is the separation constant. Unlike for Kerr with an irregular singular point at $r=\infty$, the ODE has three regular singular points located at $r=(r_+,r_-,\infty)$, and can be solved by hypergeometric functions. The corresponding monodromy eigenvalues at each horizon are defined as
\bea
\alpha_\pm = \frac{\omega - \Omega_\pm m}{2 (\pm \kappa_\pm)}
\eea
where $\kappa_\pm = \frac{r_+ - r_-}{4M r_\pm} $ is the surface gravity  and  $\Omega_{\pm}= \frac{a}{2M r_\pm}$ the angular velocity at the inner ($-$) and outer ($+$) black hole horizons.
In contrast, one can show that the wave equations for the other KEGs with $SL(2, R) \times U (1) $ hidden symmetry and  $SO(4,2)$ reduce to
\bea\label{eq:KG2}
\Bigg[\partial_r \left(\Delta_r  \partial_r \right)+  \frac{\left(2M\omega r_+  - a m\right)^2}{(r-r_+)(r_+ - r_-)}  - \frac{\left(2M k \omega r_+  - a m\right)^2}{(r-r_-)(r_+ - r_-)}  -  \hat{K}_{\ell, s} \Bigg] \hat{R}_s = 0
\eea
with $s=0$, separation constant $\hat{K}_{\ell,s}= \ell (1+\ell)$ and a deformation parameter $k=\pm1$ \footnote{Alternatively, as shown in \cite{Lowe:2011aa}, one can set $k=r_-/r_+$ such that the wave equation reduces to \eqref{eq:KG1}}. The $k$ deformation is such that the linearized equation of motion simplifies to the Schwarzschild equation as $\omega\rightarrow 0$, as demonstrated in \cite{Lowe:2011aa}. Specifically, the case $k=+1$ corresponds to the KEG with $SO(4,2)$ symmetry, while $k=-1$ corresponds to the KEG with hidden symmetry $SL(2, R) \times U (1)$.  While the location of the regular singular points remain invariant in comparison to \eqref{eq:KG1} the monodromies become
\bea
\alpha^{k=\pm 1}_+ = \frac{\omega - \Omega_+ m}{2 (+ \kappa_+)}\,,\qquad \alpha^{k=\pm1}_- = \frac{ k \omega -  \Omega_+ m}{2 (- \kappa_+ )}
\eea
Shifting the position of the monodromies in the black hole interior may a priori not seem problematic. However, as shown in \cite{Perry:2023wmm} the dynamical tidal responses for Kerr black holes will capture the low frequency responses when considering \eqref{eq:KG1}. The internal structure of the KEGs is in fact relevant to invariant quantities characterizing the black holes.

Let us conclude this section with some additional observations. Capping the inner black hole horizon region leading to \eqref{eq:KG2} does not alter the thermodynamics of the outer horizon but will impact the interpretation of the inner black hole mechanics as explored in \cite{Castro:2012av} as well as the product of areas property $A_+ A_- \in \mathbb{Z}$ depend only on the quantized charges (See \cite{Cvetic:2010mn} and references therein). 

In this context as well, the internal alterations in KEGs with $SL(2, R) \times U (1) $  and  $SO(4,2)$ hidden symmetry respect to Kerr will impact the “first law” for the inner Cauchy horizons of black holes \cite{Castro:2012av}. The areas of the inner Cauchy black hole horizon at $r=r_-$ for KEGs are
\bea
&&A_-^{SL(2,R)\times SL(2,R)}=\pi (r_{-}^{2}+a^{2})\equiv A_-^{Kerr}\,,\\
&&A_-^{SO(4,2)}= -A_-^{SL(2,R)\times U(1)}
%=\pi (r_{+}^{2}+a^{2}-4M r_+)\,,\\
=\pi (r_{+}^{2}+a^{2})\equiv A_+^{Kerr}\,.
\eea
While the inner horizon area remains invariant in the first case respect to Kerr, for the latter two cases there is a shift. Finally, we note that the monodromy matrices $M_{\gamma}$ characterized by the monodromy eigenvalues around the $r=\gamma$ singular point must obey the global trivial identity
\bea
M_{r_+}M_{r_-}M_{\infty}=1\,.
\eea
Hence, this relation may explain why the inner horizon data for each KEGs appears in black hole scattering computations. Modifying the inner information of the black hole, as in the Starobinsky and Love KEG, will result in a corresponding modification in the scattering data.
%%%%%%%%%%%%%%%%%SECTION%%%%%%%%%%%%%

\section{Quasi-normal modes}
\label{sec:QNM}

In this section we will explicitly compute the quasinormal mode (QNM) frequencies for the KEG. Following \cite{Cvetic:2013lfa}, we will argue that only the $SL(2,R)\times SL(2,R)$ KEG in \cite{Cvetic:2011dn,Cvetic:2014ina} is continuously connected to Kerr black holes.

Each KEG has a distinctive spectrum, characterized by the QNM frequencies that represent the scattering resonances of the black-hole spacetime. The QNMs correspond to specific boundary conditions where the waves are purely outgoing at infinity and purely ingoing at the horizon. They thus correspond to poles of the transmission and reflection amplitudes. 

Our starting point is the radial equation \eqref{eq:KG2}. By making the following change of coordinates, and redefinition of the functions
\bea\label{sols}
 z=\frac{r-r_+}{r-r_-} \,,\qquad \hat{R}_s(r)=(r-r_-)^p\,(r-r_+)^q\,w\,,
\eea 
we can bring the equation to the form
\bea\label{eqHyper}
z\,(1-z) \,\, \frac{d^2 w}{dz^2}+[\mathfrak{c}-(\mathfrak{a}+\mathfrak{b}+1)\,z] \, \,\frac{dw}{dz}-\mathfrak{a} \mathfrak{b} \,\, w=0\,
\eea
where we choose
\bea
&&p= \nonumber i \,\alpha_+ - (1+\ell)\,,\qquad q=- i \alpha_+ \,,\\
&&\mathfrak{a} = 1+\ell + i (\alpha_+-\alpha_-) ,\qquad \mathfrak{b} = 1+\ell - i (\alpha_+ +\alpha_-) ,\qquad \mathfrak{c} = 1+ 2 i \alpha_+\,.
\eea
A solution which is ingoing on the future horizon must be regular at the horizon $z=0$,  and this implies
\bea
\hat{R}_s(z)= c_1 (1-z)^{-(p+q)}\,z^q\,F[\mathfrak{a},\mathfrak{b},\mathfrak{c};z]\,.
\eea
where $c_1$ is a normalization constant. We can then analyze the behavior at large distances, for $z \rightarrow 1$
\bea
F[\mathfrak{a},\mathfrak{b},\mathfrak{c};z]&=&\frac{\Gamma(\mathfrak{c})\Gamma(\mathfrak{c}-\mathfrak{a}-\mathfrak{b})}{\Gamma(\mathfrak{c}-\mathfrak{a})\Gamma(\mathfrak{c}-\mathfrak{b})}F[\mathfrak{a},\mathfrak{b},\mathfrak{a}-\mathfrak{b}-\mathfrak{c}+1;1-z]\nonumber\\
 &&+(1-z)^{\mathfrak{c}-\mathfrak{a}-\mathfrak{b}}\,\frac{\Gamma(\mathfrak{c})\Gamma(\mathfrak{a}+\mathfrak{b}-\mathfrak{c})}{\Gamma(\mathfrak{a})\Gamma(\mathfrak{b})} F[\mathfrak{c}-\mathfrak{a},\mathfrak{c}-\mathfrak{b},\mathfrak{c}-\mathfrak{a}-\mathfrak{b}+1;1-z]\,.\nonumber
\eea
This vanishing of the the
solution at the boundary, $z=1$ will only be possible for a discrete set of complex frequencies
$\omega$ called QNM frequencies.
We therefore must set
\bea
\mathfrak{a}=-N_L\,,\qquad \mathfrak{b}=-N_R
\eea
where $N_{L,R}=0,1,...$.
This gives the following remarkably simple formulae for the frequencies of the QNMs
\bea\label{eq:qnm}
\omega \, (1-k) &=& - i \, 4\pi \, \nonumber T_+ (1+\ell+N_L)\,,\\
\omega\,  (1+k) &=& - i \,  4\pi \,  T_+ (1+\ell+N_R) +2 m \Omega_+
\eea
As for the Kerr black hole, there exists a discrete spectrum of QNMs. These modes consist of two branches: one is overdamped and the other is underdamped, both exhibiting rotational splitting (for $k\ne \pm 1$).
In the KEGs of interest to this paper, we argue that both damped frequencies only exist in the $SL(2,R)\times SL(2,R)$ KEG case for $k=r_-/r_+$. This is exactly what was found ten years ago by inspecting the QNMs in a curved background characterized by an \( SL(2,R) \times SL(2,R) \) symmetry in \cite{Cvetic:2014ina}. It was argued that the near-horizon geometry of Kerr black holes is accurately represented by the KEG solution, characterized by the subtracted geometry. In fact, the family of modes given by the second family of modes in \eqref{eq:qnm} have identical frequencies to those of \cite{Hod:2008se} where the resonances take the form:
\bea\label{eq:nearQNM}
\omega=- 2 i \pi T_+ (l+1+n) + m \,\Omega_+\,,
\eea
 in the $Im(\omega) \ll Re(\omega) \ll 1/M$  regime.
 This spectrum requires a Kerr black hole in the slow rotation, near-extremal limit $\sqrt{M^2-a^2}\ll a \ll M$ and $m>0$. 

The first (overdamped) branch of \eqref{eq:qnm}  corresponds to negative imaginary frequencies with absolute values significantly larger than those of the second set that was not observed in \cite{Hod:2008se}.
Let us conclude this section with some additional  comments for the deformation $k=\pm 1$ associated with the Love and Starobinsky KEG models. One could contemplate these scenarios and notice that unlike the KEG with hidden $SL(2,R) \times SL(2,R)$ symmetry, these cases feature only one branch of QNMs in each case. According to Keshet and Neitzke \cite{Keshet_2008}, a CFT description is expected to emerge in the regime of large damping, where the transmission and reflection amplitudes adopt a familiar CFT-like form. It is interesting to note that for $k = +1$ in the KEG with $SO(4,2)$ hidden symmetry, there is only one damped frequency branch, which precisely matches the expression for Kerr given by \eqref{eq:nearQNM}. Finally, we observe that when $k = -1$, the absence of overdamped resonance may suggest that the KEG with $SL(2,R) \times U(1)$ symmetry is not linked to a Kerr quasinormal spectrum and might not be relevant in the context of a dual CFT description.

%%%%%%%%%%%%%%%%%%%%%SECTION%%%%%%%%%%%%%%
\section{Separability in KEG}
\label{sec:separability}

The rationale for the separability of the Teukolsky equation \eqref{Laplacian} in the Kerr background has been understood for a considerable time now, and it is attributed to the presence of a nontrivial Exact Killing-St\"ackel Tensor (EKST) $K_{\nu\rho}$ satisfying
\bea\label{EKST}
\nabla_{(\mu}K_{\nu\rho)}=0 \; .
\eea
The properties of Kerr do not readily extend to the broader class of KEGs. Although the Kerr geometry has a nontrivial EKST satisfying \eqref{EKST} \cite{Keeler:2012mq}, other geometries within the class we are examining will only exhibit a Conformal Killing-St\"ackel Tensor (CKST), which satisfies
\bea\label{CKST}
\nabla_{\mu} M_{\nu\rho} = g_{(\mu\nu} N_{\rho)}
\eea
with nonzero $N_{\rho}$. See e.g. \cite{Cvetic:2011dn} for a review on the classification of Killing tensors. In these geometries, $T^{\mu}T^{\nu} M_{\mu\nu}$ is only conserved along affinely parameterized null geodesics. The existence
of a nontrivial CKST ensures separability of the massless
scalar wave equation or the Hamilton-Jacobi equation. In contrast, the separability of the massive equation is only guaranteed
by an exact Killing-St\"ackel tensor solving \eqref{EKST}. We will see that in certain cases such a tensor can be found for KEGs.

The CKST for the KEG  metric \eqref{ACmetric} satisfy the condition \eqref{CKST} with the tensors
\begin{align}
P^{\mu\nu}(\theta)\partial_{\mu}\partial_{\nu} & = \partial_{\theta}^{2}+\frac{1}{\sin^{2}\theta}\partial_{\phi}^{2}-\varrho_{\theta}\partial_{t}^{2}+2a \partial_{t}\partial_{\phi} \\
S^{\mu\nu}(r)\partial_{\mu}\partial_{\nu} &= \Delta \partial^{2}_{r}-\varrho_{R}\partial_{t}^{2} - \frac{1}{\Delta}(\mathcal{A}_{red}\partial_{t}+a \partial_{\phi})^{2}-2a\partial_{t}\partial_{\phi}.
\end{align}
where $\mathcal{A}_{red}=A_0\,  G/( a \sin^2 \theta ) $ and
\bea
\varrho \equiv \frac{\Delta_0 - \mathcal{A}_{red}^2}{G_0}=
 \varrho_{\theta}(\theta) + \varrho_{R} (r)
\eea

Thus \(P^{\mu\nu}\) and \(S^{\mu\nu}\) are CKSTs with inhomogeneous terms respectively given by
\begin{equation}
V^\mu = \frac{\partial_\theta(\Delta_0^{1/2})}{\Delta_0^{1/2}} \delta^\mu_\theta,    
\qquad
U^\mu = \frac{\partial_r(\Delta_0^{1/2}) \Delta}{\Delta_0^{1/2}} \delta^\mu_r.
\end{equation}
The CKSTs constructed can in certain cases be promoted to EKST satisfying \eqref{EKST}.
\bea
K_{\mu\nu}(r, \theta) = \frac{1}{\Delta_0^{1/2}} [h(r)P_{\mu\nu}(\theta) - f(\theta)S_{\mu\nu}(r)]
\eea
where the functions $h,f$ are defined by
\bea
\Delta_0(r, \theta)^{1/2} = f(\theta) + h(r)
\eea 
One of the KEGs of interest was constructed in \cite{Cvetic:2011dn}, by demanding that they promote approximate $SL(2,R)\times SL(2,R)$ near horizon symmetries to exact symmetries. Despite the enhanced symmetry, as shown in \cite{Keeler:2012mq}, the CKSTs of these geometries cannot be promoted to EKSTs. 

Interestingly, we find that for other KEGs, such as the one exhibiting an $SO(4,2)$ or $SL(2,R)\times U(1)$ hidden symmetry, one can promote the CKST to EKST. Collecting formulae, we finally find the EKST for these cases:
\bea
K_{SL(2,R)\times U(1)}^{\mu\nu}&=& \frac{[h(r)P_{\mu\nu}(\theta) - f(\theta)S_{\mu\nu}(r)]}{(r_+^2+a^2)(1+\beta^2\Omega^2 \sin^2\theta)^{1/2}} \,,
\eea
where
\bea
f(\theta)+ h(r) = (r_+^2+a^2)(1+\beta^2\Omega^2 \sin^2\theta)^{1/2}\,\qquad \varrho_{\theta} + \varrho_{R} =(r_+^2+a^2)^2\beta^2 \Omega^2 /a^2
\eea
and 
\bea
K_{SO(4,2)}^{\mu\nu}&=& \frac{[h(r)P_{\mu\nu}(\theta) - f(\theta)S_{\mu\nu}(r)]}{(r_+^2+a^2)}\,.
\eea
where
\bea
f(\theta)+ h(r) = (r_+^2+a^2)\,\qquad \varrho_{\theta} + \varrho_{R}=0
\eea
These expressions can be recast in many equivalent ways by freely choose the separation constants between the $f,h$ and $\varrho_{\theta}, \varrho_{R}$.
%\textcolor{red}{MJR:  What is their relation to SL(2,R) symmetry?}.

The existence of EKST suggests that both of these geometries satisfy the strongest form of separability of the Teukolsky equation, known as {\it quantum separability}. This refers to the situation where 
\bea
[\nabla_\mu K_{\mu\nu} \nabla_\nu, g_{\lambda\sigma} \nabla_\lambda \nabla_\sigma] = 0
\eea
This condition does not differentiate between massive and massless Klein-Gordon equation, and it evidently implies the separability of the geodesic equations.

%%%%%%%%%%%%%%%%%SECTION%%%%%%%%%%%%%
\section{Discussion}
\label{sec:Disc}

We have identified a matter configuration that supports these geometries through a rescaling limit. Consequently, we argue that the effective field theory description of Kerr black holes is encoded within the black holes of the STU model rather than vacuum General Relativity. The scaling solution generating technique that we detailed could be used for a wider class of stationary solutions, beyond the KEG solutions that we used. While the application of this method was known for the subtracted geometries \cite{Cvetic:2011dn}, two new examples, the Love and Starobinsky KEGs, were obtained via this method. Under the assumption of static configurations, we have utilized the Harrison transformations to embed the static KEG within the supergravity theories. 

Our quantitative analysis compares physical quantities, internal structures, and geodesic equations across different effective Kerr spacetimes. While their thermodynamic properties align with those of Kerr, including entropy calculations, a closer examination of their internal structures revealed substantial differences. These differences point to distinct microscopic configurations and dynamics within these spacetimes, despite their macroscopic similarities.

We have given the quasinormal spectra of the KEGs, which provides further insight into the behavior of these spacetimes under linear scalar perturbations. We showed that the fundamental resonances can be expressed in terms of the black hole temperature and angular velocity. We noted the property that the generic KEGs are not isospectral to Kerr. This analysis helps to further contrast the scope of the various KEGs as a Kerr black hole. In this context, one interesting aspect that we will leave for future exploration is study of the stability of the KEGs, presenting an intriguing avenue of research.

We discussed the separability of the massless Klein–Gordon equations for KEGs of interest, those with $SO(4,2)$ or $SL(2,R)\times U(1)$ hidden symmetry, but a comprehensive examination of all KEGs remains to be done.
We showed that these KEGs admit an Exact Killing-St\"ackel Tensor. The existence of this tensor indicates that both of these geometries satisfy the strongest form of separability of the Teukolsky equation, known as quantum separability.  We can further deduce that the Klein-Gordon equation, for both massive and massless cases, is separable, and implies the separability of the geodesic equations.

We studied the thermodynamical properties of a class of KEGs on the black hole horizons. These geometries have very different asymptotic structure compared to the asymptotically flat and asymptotically AdS case. The energy density of such metrics typically falls off as inverse  of the radial distance and thus the geometry cannot have a finite total energy. Within the KEG family, only for the $SL(2,R)\times SL(2,R)$ metric explicit expressions for the mass and angular momentum are known \cite{Cvetic:2014nta}. For generic KEGs the corresponding asymptotic charges, the mass and angular momentum have not been explored in detail. New insights there would clarify the first law of thermodynamics both at the outer and inner horizon and that the Smarr formula holds.\\

%%%%%%%%%%%%%%%%%ACKNOWLEDGMENTS%%%%%%%%%%%%%
%\section*{Acknowledgements} 

{\bf\Large Acknowledgements} \\

We would like to thank Austin Joyce, Malcolm Perry, Chris Pope, Luca Santoni, Adam Solomon and Chiara Toldo for discussions. We gratefully thank the Mitchell Family Foundation at Cook's Branch workshop and the Centro de Ciencias de Benasque Pedro Pascual for their warm hospitality. MC is supported by the  by DOE Award (HEP) DE-SC0013528, the Simons Foundation Collaboration grant $\#$724069, Slovenian Research Agency (ARRS No. P1-0306) and Fay R. and Eugene L. Langberg Endowed Chair funds. NHR was supported by RYC-2016-21159 and CNS2022-135880. MJR was partially supported through the NSF grant PHY-2309270, CNS2022-135880, CEX2020-001007-S and PID2021-123017NB-I00, funded by MCIN/AEI/10.13039/501100011033 and by ERDF A way of making Europe. OV is supported by the NSF grant 2310223 and, partially, by CEX2020-001007-S and PID2021-123017NB-I00, funded by MCIN/AEI/10.13039/501100011033 and by ERDF A way of making Europe.

%%%%%%%%%%%%%%%%% Appendix%%%%%%%%%%%%%
\appendix

\section{STU black holes and subtracted geometries}
\label{app:A}

\subsection{Asymptotically flat black hole solution}

Below is a summarized description of the general scaling procedure for this type of effective Kerr solutions, as described in \cite{Chong:2004na, Cvetic:1996kv}. For a detailed explanation, which includes the reduction of the Kerr metric to 3 dimensions and its subsequent conversion back to 4 dimensions using the Kaluza-Klein reduction rules, please refer to the mentioned article. After this process, the resulting metric for the 4-dimensional rotating black hole with 4 charges is
\begin{equation}
    ds_{4}^{2}= - \frac{\rho^{2}-2mr}{W}(dt+\mathcal{B}_{(1)})^{2} + W\left(\frac{dr^{2}}{\Delta}+d\theta^{2}+\frac{\Delta\sin^{2}\theta d\phi^{2}}{\rho^{2}-2mr}\right),
\end{equation}
where $ B_{(1)}$ is the Kaluza-Klein vector from the reduction and $\rho$ and $\Delta$ are the original functions from the Kerr metric,

\begin{align}
    B_{(1)}=\nonumber\hspace{0.1cm} & \frac{2m(a^{2}-u^{2})(r c_{123}-(r-2m)s_{1234})}{a(\rho^{2}-2mr)}\text{d}\phi\\
    \rho^{2}\hspace{0.1cm}=&r^{2}+a^{2}\cos^{2}\theta \hspace{2cm}\Delta\hspace{0.1cm}=r^{2}-2mr+a^{2},
\end{align}
and the remaining elements are defined as
\begin{align}
W^{2}=\nonumber\hspace{0.1cm} & r_{1}\hspace{0.1cm}r_{2}\hspace{0.1cm}r_{3}\hspace{0.1cm}r_{4}+u^{4}+u^{2}\{2r^{2}+2mr\hspace{0.1cm}(s_{1}^{2}+s_{2}^{2}+s_{3}^{2}+s_{4}^{2})\\
& \nonumber+8m^{2}\hspace{0.1cm}c_{1234}\hspace{0.1cm}s_{1234}- 4m^{2}(s_{123}^{2}+s_{124}^{2}+s_{134}^{2}+s_{234}^{2}+ 2 s_{1234}^{2}\}\\
s_{i_{1}\cdot i_{n}}= \hspace{0.1cm}&\sinh{\delta_{i_{1}}}\cdot\sinh{\delta_{i_{n}}}\hspace{1cm} c_{i_{1}\cdot i_{n}}= \hspace{0.1cm}\cosh{\delta_{i_{1}}}\cdot\cosh{\delta_{i_{n}}}\hspace{1cm} u= a \hspace{0.1cm} \cos\theta,
\end{align}
with $\delta_{i}$ being the charge parameter. 

In this context, the 4-dimensional solution for 4 charges consists of 4 gauge potentials, with 2 of them associated with the electric charges, \begin{align}
\label{electric}
    A_{12} =\nonumber \frac{2m}{aW^2} \{( & r_1 \hspace{0.1cm} r_3 \hspace{0.1cm}  r_4 + r u^2 ) \left[ c_{2}\hspace{0.1cm}  s_{2}\hspace{0.1cm}  a {dt} - (a^2 - u^2) ( c_{134}\hspace{0.1cm}  s_2 - s_{134}\hspace{0.1cm}  c_2 )\hspace{0.1cm} d\phi \right] \\
    & + 2m u^2\left[e_2 \nonumber\hspace{0.1cm}  a dt - (a^2 - u^2) \hspace{0.1cm}  s_{134} \hspace{0.1cm}  c_2 \hspace{0.1cm} d\phi\right]\}\\
    \mathcal{A}_{1}^{2} =\nonumber \frac{2m}{aW^2} \{( & r_1 \hspace{0.1cm} r_2 \hspace{0.1cm}  r_3 + r u^2 ) \left[ c_{4} \hspace{0.1cm}  s_{4} \hspace{0.1cm} a dt - (a^2 - u^2) ( c_{123} \hspace{0.1cm}  s_4 - s_{123} \hspace{0.1cm}  c_4 ) \hspace{0.1cm} d\phi \right] \\
    & + 2m u^2\left[e_4 \hspace{0.1cm} a dt - (a^2 - u^2) s_{123}\hspace{0.1cm}  c_4 \hspace{0.1cm} d \phi\right]\},
\end{align}
and the other 2, analytically more complex, associated with the magnetic charges, 
\begin{align}
\label{magnetic}
    A_{(1)1} = \nonumber \frac{2m u}{aW^2} \{( & r\hspace{0.1cm}  r_1 +u^2 ) \left[( c_{234}\hspace{0.1cm} s_{1}-s_{234}\hspace{0.1cm} c_{1})\hspace{0.1cm}  a \text{dt} - c_{1}\hspace{0.1cm} s_{1}\hspace{0.1cm} a^{2}\text{d}\phi \right]+2m\hspace{0.1cm}  r_{1}\hspace{0.1cm}  s_{234}\hspace{0.1cm} c_{1}\hspace{0.1cm} a \text{dt} \\
    &-(c_{1} \nonumber\hspace{0.1cm}s_{1}\hspace{0.1cm}(r_{1}\hspace{0.1cm}r_{2}\hspace{0.1cm}r_{3}\hspace{0.1cm}r_{4} + u^{2}\left[r^{2}+2mr\hspace{0.1cm}(s_{2}^{2}+s_{3}^{2}+s_{4}^{2})-4m^{2}\hspace{0.1cm}s_{234}^{2}\right])\\
    & \nonumber+4m^{2}u^{2}\hspace{0.1cm}c_{234}\hspace{0.1cm}s_{234}\hspace{0.1cm}s_{1}^{2}+2m\hspace{0.1cm}e_{1}\hspace{0.1cm}(a^{2}\hspace{0.1cm}r_{1}-r\hspace{0.1cm}u^{2}))\text{d}\phi\},\\
    \mathcal{A}_{(1)1} = \nonumber\frac{2m u}{aW^2} \{( & r\hspace{0.1cm}  r_3 +u^2 ) \left[( c_{124}\hspace{0.1cm} s_{3}-s_{124}\hspace{0.1cm} c_{3})\hspace{0.1cm}  a \text{dt} - c_{3}\hspace{0.1cm} s_{3}\hspace{0.1cm} a^{2}\text{d}\phi \right]+2m\hspace{0.1cm}  r_{3}\hspace{0.1cm}  s_{124}\hspace{0.1cm} c_{3}\hspace{0.1cm} a \text{dt} \\
    &-(c_{3} \nonumber\hspace{0.1cm}s_{3}\hspace{0.1cm}(r_{1}\hspace{0.1cm}r_{2}\hspace{0.1cm}r_{3}\hspace{0.1cm}r_{4} + u^{2}\left[r^{2}+2mr\hspace{0.1cm}(s_{1}^{2}+s_{2}^{2}+s_{4}^{2})-4m^{2}\hspace{0.1cm}s_{124}^{2}\right])\\
    & +4m^{2}u^{2}\hspace{0.1cm}c_{124}\hspace{0.1cm}s_{124}\hspace{0.1cm}s_{3}^{2}+2m\hspace{0.1cm}e_{3}\hspace{0.1cm}(a^{2}\hspace{0.1cm}r_{3}-r\hspace{0.1cm}u^{2}))\text{d}\phi\},
\end{align}
where $ r_{i}= r+2ms_{i}^{2} $ in the solutions shown above. Likewise, the expressions for the $ e_{i} $ are defined as follows,
\begin{align}
    e_{1}=\nonumber& c_{234}\hspace{0.1cm}s_{234}(c_{1}^{2}+s_{1}^{2})-c_{1}\hspace{0.1cm}s_{1}(s_{23}^{2}+s_{24}^{2}+s_{34}^{2}+2s_{234}^{2})\\
    e_{2}=\nonumber& c_{134}\hspace{0.1cm}s_{134}(c_{2}^{2}+s_{2}^{2})-c_{2}\hspace{0.1cm}s_{2}(s_{13}^{2}+s_{14}^{2}+s_{34}^{2}+2s_{134}^{2})\\
    e_{3}=\nonumber& c_{124}\hspace{0.1cm}s_{124}(c_{3}^{2}+s_{3}^{2})-c_{3}\hspace{0.1cm}s_{3}(s_{12}^{2}+s_{24}^{2}+s_{14}^{2}+2s_{124}^{2})\\
    e_{4}=& c_{123}\hspace{0.1cm}s_{123}(c_{4}^{2}+s_{4}^{2})-c_{4}\hspace{0.1cm}s_{4}(s_{23}^{2}+s_{12}^{2}+s_{13}^{2}+2s_{123}^{2}).
\end{align}
On the other hand, The remaining 4-dimensional fields, namely, the three dilatons $(\varphi_{1}, \varphi_{2}, \varphi_{3})$ and the three axions $(\chi_{1},\chi_{2},\chi_{3})$, are determined by the following expressions,
 
\begin{align}
    \chi_{1} = & \frac{2mu (c_{13}s_{24}-c_{24}s_{13})}{r_{1}r_{3}+u^{2}}, \hspace{1cm} \chi_{2}=\frac{2mu(c_{14}s_{23}-c_{23}s_{14})}{r_{2}r_{3}+u^{2}}\\
    \chi_{3}=& \frac{2mu(c_{12}s_{34}-c_{34}s_{12})}{r_{1}r_{2}+u^{2}}, \hspace{0.90cm} e^{\varphi_{1}}=\frac{r_{1}r_{3}+u^{2}}{W}\\
    e^{\varphi_{2}}= & \frac{r_{2}r_{3}+u^{2}}{W}, \hspace{3.1cm}e^{\varphi_{3}}=\frac{r_{1}r_{2}+u^{2}}{W}.
\end{align}

\subsection{Scaling procedure for KEG with \texorpdfstring{$SL(2,R)\times SL(2,R)$}{SL(2,R) X SL(2,R)} symmetry}
\label{SL2R}

The results presented above in the limit $\delta_{1}=\delta_{2}=\delta_{3}=\delta_{4}$ lead to the dyonic Kerr-Newman solution; however, in this subsection, another arrangement of the charges is discussed as proposed in \cite{Cvetic:2014ina}.

\subsubsection{Static case }

If a symmetry of type $SL(2,R)\times SL(2,R)$ exists, we will experience variations in the charges, potentials, and parameters used in Section \ref{sec:Matter}. In this case, we will employ  the fields defined in equations \eqref{electric} and \eqref{magnetic}
%$* F_1= F_2= * {\cal F} _1\equiv F\, $  and  $ {\cal F}_2\equiv {\cal F}$, with corresponding potentials $A_{1}=A_{2}=A_{3}=A$ and $A_{4}=\mathcal{A}$ 
and the choice of charge parameters $\bar{\delta}_{1}=\bar{\delta}_{2}=\bar{\delta}_{3}=\bar{\delta}$ and $\bar{\delta}_{4}=\bar{\delta}_{0}$. The limit is
implemented by means of the following scalings:
\bea
&&\bar{r}= r  \epsilon,   \quad \bar{t}= t{\epsilon^{-1}},  \quad  \bar{m}=  m \epsilon\, , \ \cr
&&2 \bar{m} \sinh^2 \bar{\delta} \equiv Q = {2m}{\epsilon^{-1/3}} (\Pi_c^2-\Pi_s^2)^{1/3},   \quad \sinh^2 \bar{\delta}_0=\frac {\Pi_s^2}{\Pi_c^2-\Pi_s^2}\, ,   \
\label{scalings} \eea
Considering $\epsilon\rightarrow 0$, the gauge potentials and gauge fields related to the metric take the following form,
\bea
&&\chi_1=\chi_2=\chi_3=0 , \ \ e^{\varphi_1} =e^{\varphi_2} =e^{\varphi_3} = \frac{Q^2} {\Delta_s^{\frac{1}{2}}}, \cr
&&A=-\frac{r}{Q}\, dt , \ \ {\cal A}= \frac{Q^3 (2m) \Pi_c\Pi_s }{(\Pi_c^2-\Pi_s^2)\Delta_s}\, dt ,
\eea
where the corresponding field strengths are 
\bea&&F_{t\, r}= \frac{1}{Q}, \quad {\cal F}_{t\, r}= \frac{Q^3(2m)^4  \Pi_c\Pi_s }{\Delta_s^2}\, .
\eea
where
\beq
\Delta_{0s}\to \Delta_s= (2m)^3 r (\Pi_c^2-\Pi_s^2) + (2m)^4 \Pi^2_s\ .
\eeq 
It is important to note that the static case has no axions.

\subsubsection{Rotating case}

Now let us perform the same calculation taking into account rotation, which means that the spin parameter is not zero, i.e., $a\neq 0$. Under these assumptions, the result obtained for the scalar fields associated with the metric is,
\begin{align*}
&\nonumber \chi_1=\chi_2=\chi_3=-\frac{2ma (\Pi_{c}-\Pi_{s}) \cos{\theta}}{Q^{2}},
&e^{\varphi_1} =e^{\varphi_2} =e^{\varphi_3} = \frac{Q^2} {\Delta^{1/2}},
\end{align*}
where the appearance of axions, which are nonexistent in the static solution, is highlighted. On the other hand, the gauge fields of this solution correspond to
\bea
\nonumber &&A=-\frac{r}{Q}dt+\frac{(2m)^2 a^2[2m\Pi_s^2 -r (\Pi_c-\Pi_s)^2]\cos^ 2\theta}{Q\Delta}dt \cr
&&-\frac{2m\, a(\Pi_c-\Pi_s)\sin^2\theta }{Q}\left(1+\frac{(2m)^2a^2(\Pi_c-\Pi_s)^2\cos^2\theta}{\Delta}\right)d\phi\, ,
\cr
\nonumber &&{\cal A}= \frac{ Q^3[(2m)^2 \Pi_c\Pi_s + a^2 (\Pi_c-\Pi_s)^2\cos^2\theta]}{2m(\Pi_c^2-\Pi_s^2)\Delta}\, dt \, + \frac{Q^32m\,a(\Pi_c-\Pi_s)\sin^2\theta}{\Delta}   \, d\phi.
\eea

\section{Harrison Transformation}
\label{app:B}

Given a solution of four-dimensional Einstein-Maxwell dilation theory of the form
\begin{equation}
    ds^{2}=-e^{2U}dt^{2}+e^{-2U}\gamma_{ij}dx^{i}dx^{j}, \hspace{2cm}F_{i0}=\partial_{i}\Psi, \hspace{1cm}e^{-2\varphi}
\end{equation}
We can derive an effective three-dimensional action by performing a dimensional reduction along a timelike Killing vector instead of a spacelike one. This approach allows us to define
\begin{equation}
P=e^{(\alpha-1)U}e^{-(\alpha+1)\varphi}\left (
  \begin{array}{cc}
    e^{2(U+\varphi \alpha)}-(1+\alpha^{2})\Psi^{2} & -\sqrt{1+\alpha^{2}}\Psi \\
     \sqrt{1+\alpha^{2}}\Psi & -1 
  \end{array}
\right )
\end{equation}
and we can express the Lagrangian density, which is invariant under $GL(2;\mathbb{R})$, in the following form
\begin{equation}
\mathcal{L}_{3}=\mathcal{R}+\frac{1}{1+\alpha^{2}}\gamma_{ij}\textrm{Tr}(\partial_{i}P\partial_{j}P^{-1}).
\end{equation}

By applying a specific element of SO(1; 1), corresponding to a Harrison transformation 
\begin{equation}
H=\left (
  \begin{array}{cc}
    1 & 0 \\
     \beta & 1 
  \end{array}
\right )
\end{equation}
on $P$, we can find $P^{'}=HPH^{-1}$ with new potentials, and hence a new four-dimensional solution 
\begin{equation}
    ds^{2}_{new}=-e^{2U'}dt^{2}+e^{-2U'}\gamma_{ij}dx^{i}dx^{j}, \hspace{2cm}F_{i0}^{'}=\partial_{i}\Psi^{'}, \hspace{1cm}e^{-2\varphi'},
\end{equation}
where
\begin{align}
  e^{2U'} & =\Lambda^{\frac{-2}{(1+\alpha^{2}}}e^{2U}, \hspace{1.2cm}e^{-2\varphi'} =\Lambda^{\frac{-2\alpha}{(1+\alpha^{2}}}e^{-2\varphi}\\
  \Psi^{'}&=\Lambda^{-1}\left(\Psi +\frac{\beta(e^{2(U+\alpha\varphi)}-(1+\alpha^{2})\Psi^{2})}{\sqrt{1+\alpha^{2}}}\right)\\
  \Lambda & = (\beta\psi +1)^{2}-\beta^{2}e^{2}(U+\alpha \varphi).
\end{align}

If we set $\alpha=0$, we obtain the expected result for the Love and Starobinsky metrics in the spinless limit, $a\rightarrow 0$. The initial solution is 
\begin{equation} 
 \textrm{Schwar} = \left\{
        \begin{array}{ll}
            e^{2U} &\hspace{-0.1cm} =1 - \frac{2M}{r}, \hspace{1cm}  \alpha= 0 \\
             \varphi &\hspace{-0.1cm}=0, \hspace{2cm}\Psi=0 
        \end{array}
        \right.
\end{equation}
and after the transformation the new solution is characterized by
\begin{equation} 
 \textrm{Schwar New} = \left\{
        \begin{array}{ll}
            e^{2U'} &\hspace{-0.1cm} =-\frac{r(2M-r)}{4 M^{2}}, \hspace{0.5cm}  \alpha= 0, \hspace{0.5cm} \Lambda=\frac{2M}{r}\\
             \varphi^{'} &\hspace{-0.1cm}=0, \hspace{1.77cm}\Psi^{'}=-1+\frac{r}{2M} 
        \end{array}
        \right.
\end{equation}

%%%%%%%%%%%%%%%%% bibliography%%%%%%%%%%%%%

\addcontentsline{toc}{section}{References}

\bibliography{Effective.bib}

\end{document}